\documentclass[sigconf]{acmart}

\usepackage[utf8]{inputenc} 
\usepackage[T1]{fontenc}    
\usepackage{hyperref}       
\usepackage{url}            
\usepackage{booktabs}       
\usepackage{amsfonts}       
\usepackage{nicefrac}       
\usepackage{microtype}      
\usepackage{lipsum}
\usepackage{graphicx}
\usepackage{listings}
\usepackage{siunitx}

\AtBeginDocument{%
  \providecommand\BibTeX{{%
    \normalfont B\kern-0.5em{\scshape i\kern-0.25em b}\kern-0.8em\TeX}}}

\copyrightyear{2023} 
\setcopyright{rightsretained} 

\definecolor{shadecolor}{RGB}{255,100,100}
\iffalse
    \newcommand{\tofix}[1]{\colorbox{shadecolor}{#1}}
    \newcommand{\say}[2]{{\color{shadecolor} \textbf{#1: }#2}}
\else
    \newcommand{\tofix}[1]{}
    \newcommand{\say}[2]{}
\fi

\begin{document}

\title{MPI Application Binary Interface Standardization}

\author{Jeff R. Hammond}
\affiliation{
  \institution{NVIDIA Helsinki Oy }
  \city{Helsinki}
  \country{Finland}
  \postcode{00180}
}
\email{jeffpapers@nvidia.com}
\orcid{0000-0003-3181-8190}

\author{Lisandro Dalcin}
\affiliation{
  \institution{Extreme Computing Research Center}
  \institution{KAUST}
  \city{Thuwal}
  \country{Saudi Arabia}}
\email{dalcinl@gmail.com}
\orcid{0000-0001-8086-0155}

\author{Erik Schnetter}
\affiliation{
  \institution{Perimeter Institute for Theoretical Physics}
  \city{Waterloo}
  \state{Ontario}
  \country{Canada}
}
\email{eschnetter@perimeterinstitute.ca}
\orcid{0000-0002-4518-9017}
  
\author{Marc Pérache}
\affiliation{
  \institution{CEA DAM}
  \streetaddress{F-91297}
  \city{Arpajon}
  \country{France}
  \postcode{F-91297}
}
\email{marc.perache@cea.fr}
\orcid{0000-0003-1615-2749}
  
\author{Jean-Baptiste Besnard}
\affiliation{
  \institution{ParaTools}
  \city{Bruy\`{e}res-le-Ch\^{a}tel}
  \country{France}
}
\email{jbbesnard@paratools.fr}
\orcid{0000-0001-6500-6786}

\author{Jed Brown}
\affiliation{
  \institution{University of Colorado Boulder}
  \city{Boulder}
  \state{Colorado}
  \country{USA}
}
\email{jed@jedbrown.org}
\orcid{https://orcid.org/0000-0002-9945-0639}

\author{Gonzalo Brito Gadeschi}
\affiliation{
  \institution{NVIDIA GmbH}
  \city{Munich}
  \country{Germany}
}
\email{gonzalob@nvidia.com}
\orcid{https://orcid.org/0000-0003-1138-3679}

\author{Joseph Schuchart}
\affiliation{
  \institution{University of Tennessee, Knoxville}
  \city{Knoxville}
  \state{Tennessee}
  \country{USA}
}
\email{schuchart@utk.edu}
\orcid{https://orcid.org/0000-0003-2041-7877}

\author{Simon Byrne}
\affiliation{
  \institution{California Institute of Technology}
  \city{Pasadena}
  \state{California}
  \country{USA}
}
\email{simonbyrne@caltech.edu}
\orcid{https://orcid.org/0000-0001-8048-6810}

\author{Hui Zhou}
\affiliation{
  \institution{Argonne National Laboratory}
  \city{Lemont}
  \state{Illinois}
  \country{USA}
}
\email{zhouh@anl.gov}
\orcid{https://orcid.org/0000-0002-4422-2911}

\renewcommand{\shortauthors}{Hammond, et al.}

\begin{abstract}
MPI is the most widely used interface for high-performance computing (HPC) workloads. 
Its success lies in its embrace of libraries and ability to evolve while maintaining 
backward compatibility for older codes, enabling them to run on new architectures for many years. 
In this paper, we propose a new level of MPI compatibility: 
a standard Application Binary Interface (ABI).
We review the history of MPI implementation ABIs,
identify the constraints from the MPI standard and ISO C,
and summarize recent efforts to develop a standard ABI for MPI. 
We provide the current proposal from the MPI Forum's ABI working group, 
which has been prototyped both within MPICH and as an independent abstraction layer called Mukautuva. 
We also list several use cases that would benefit from the definition of an ABI 
while outlining the remaining constraints.
\end{abstract}



\received{15 May 2023}

\maketitle


\section{Introduction}

MPI \cite{mpiforum:4-0} has always been an Application Programming Interface (API) standard,
which means that it is standardized in terms of the C and Fortran programming languages.
Implementations are not constrained 
in how they define opaque types 
(for example, \texttt{MPI\_Comm}), 
which means they compile into different binary representations. 
This is fine for users who only use one implementation, 
or are content to recompile their software for each of these. 
Many users, including those building both traditional C/C++/Fortran libraries 
and new languages that use MPI via the C ABI, are tired of the duplication of effort 
required because MPI lacks a standard Application Binary Interface (ABI).

The potential for implementation agnosticism~\cite{Gropp:ABI,MPI-Adapter} and 
specifically an ABI~\cite{Lindahl:ABI}, has been recognized for many years.
However, no serious effort was made to standardize an ABI, for a variety of reasons.
Some of the forces acting against ABI standardization were the diversity of HPC
systems, the prevalence of static linking, and the lack of adoption of third-party languages.
Over the past 20 years, the HPC hardware and software ecosystem has changed dramatically.
\say{SB}{Could you describe how it has changed? I haven't been in HPC that long, so can't really say for sure, but my impression is that there has been an increased dominance of the two major open-source implementations (MPICH and Open MPI), either directly or via derived implementations.}
Distributing software packages through shared libraries is now common.
Package managers, including HPC-oriented ones such as Spack~\cite{gamblin2015spack}, distribute binaries that depend on MPI.
There is increasing adoption of MPI by applications written in languages 
other than C and Fortran~\cite{rust4hpc, mpi4py, MPI.jl}.
The MPICH ABI Initiative~\cite{supalov201420,MPICH-ABI-Initiative} was the first serious effort 
to create mutually interoperable MPI implementations, 
by reconciling small differences between the ABIs of MPICH and MPICH-based implementations. 
This allows applications compiled against appropriate versions of  MPICH, Intel MPI, Cray MPI, MVAPICH2 
and other implementations to run using the shared libraries from of any of the other implementations. 
This is especially useful to leverage the level of platform-specific specialization 
that goes into some of these libraries.

Since 2014, the appetite for MPI implementation compatibility has grown dramatically for at least two reasons. 
First, containers are an increasingly popular mechanism for distributing HPC software. 
Singularity~\cite{DKPanda-containers,Veiga-containers} and Shifter~\cite{benedicic2017portable,Shifter-MPI}, 
among others, now allow complex scientific applications to be shared 
more easily by packing them as self-sustained software images. 
However, container portability is hindered~\cite{Hursey-containers} by both 
the lack of a common launch methodology\footnote{
    Note the PMIx standard~\cite{castain2017pmix} has made important progress on addressing this issue.
} 
and the absence of an MPI ABI -- preventing the advent of portable containers featuring MPI programs.
Second, MPI is now used by applications written in languages like Python, Julia, and Rust, 
which are currently required to build and test against all supported implementations and 
support the end-user installation of their MPI support against the implementation of the user's system. 
A standard ABI would eliminate the $O(N)$ cost of packaging and simplify testing. 
The $O(N)$ costs due to MPI implementation ABIs are not unique to these languages.

In the rest of this paper, we describe 
the constraints associated with an MPI ABI,
the potential benefits for the HPC ecosystem,
and
the proposed ABI implementation as defined by the ABI working group,
Performance experiments demonstrate that a high-quality implementation
of the standard ABI in MPICH has negligible overhead, while
the third-party implementation in Mukautuva has a tolerable overhead.
We also discuss important considerations for compatibility besides
the C ABI, including library naming, launchers, and Fortran.

\say{SB}{
This makes it significantly easier to provide pre-compiled binary dependencies. e.g. the Julia package manager provides a default MPI implementation for a given system, and also downstream binaries built against it (e.g. PETSc, P4est, AMReX, ADIOS2, and soon HDF5). This is a significant boost to usability, especially for the long-tail of users on lower-end systems, or developing locally.}

\say{SB}{One other frustration is a lack of any mechanism to identify what is the ABI? The closest was \texttt{MPI\_Get\_library\_version}, but it requires \texttt{MPI\_MAX\_LIBRARY\_VERSION\_STRING}, which is itself part of the ABI.}
\section{Background and Related Work}

The HPC user community has been actively working to address the issue of ABI compatibility in MPI implementations. 
For a long time, the requirements associated with ABI compatibility in MPI have led to complexities 
in terms of software deployment, particularly in large computing centers. 


Wi4MPI is a wrapper interface that implements ABI interoperability for MPI, 
supporting both Fortran and C languages~\cite{Wi4MPI-paper}. 
It can be used in two ways. 
First, users can compile their applications against  the generic MPI interface 
from Wi4MPI and then redirect them to their implementation of choice. 
Alternatively, they can redirect one implementation to another. 
For Wi4MPI to work, its wrapper interface needs to be compiled for each source and target MPI. 
The performance overhead has been shown to be minimal, making it an effective tool 
for running containers in a portable manner. 
Wi4MPI is leveraged in the Extreme-scale Scientific Software Stack~\cite{heroux2020e4s} 
as a support tool in the \texttt{e4s-cl} container launcher tool, which implements on-the-fly 
MPI detection and library translation at container launch time~\cite{skutnik2021e4s}.



A similar effort to Wi4MPI was undertaken at the Perimeter Institute for Theoretical Physics, 
leading to MPItrampoline \cite{MPItrampoline}, which defines its own ABI enabling applications' portability on several MPI runtimes.

It is known that a patent~\cite{supalov2011using} exists for a specific method of interoperating 
different MPI ABIs, preventing its use by the open-source community. 

In general, the availability of a standard ABI will simplify the tasks of these converters. 
Instead of having to implement conversions between the two APIs, 
these adaptation layers will primarily focus on compilation-related tasks, 
such as fixing dependency detection and enabling the replacement of one MPI with another.

\say{SB}{I think it is worth mentioning that a key benefit of having an ABI is that MPI wrappers like MPITrampoline can then be implemented directly via PLT trampolines, without needing to do any internal conversion themselves, reducing (potentially eliminating?) any overhead.}
\say{JB}{Tried something in this direction}

After the MPI ABI working group was formed, two efforts were started to prototype
the proposed designs, to understand their feasibility.
The first of these was Mukautuva~\cite{mukautuva-github},
which is a standalone ABI abstraction layer that maps from its own ABI
(i.e. an approximation to the one under discussion in the working group)
to MPICH and Open~MPI by redirecting MPI symbols through a translation
layer to the underlying MPI implementation, with renamed symbols 
(via \texttt{dlsym}) to avoid conflicts.
The final design of Mukautuva is unintentionally quite similar to MPItrampoline;
this convergence may be an indication of the suitability of their design.
Meanwhile, a prototype was developed in MPICH~\cite{mpich-abi-github}.
Working together, these efforts revealed the relative ease of implementing
the ABI proposal both internally and externally to an existing implementation.
They also exposed non-portable assumptions in various MPI test suites.
\section{Current ABI Designs}\label{sec:current-abi}

There are multiple aspects to an MPI ABI. Here are a few:
\begin{enumerate}
    \item The integral types of \texttt{MPI\_Aint}, \texttt{MPI\_Offset}, and  \texttt{MPI\_Count}.
    \item The \texttt{MPI\_Status} object. 
          This is a C \texttt{struct} with three standard members as well as hidden fields used by the implementation. 
    \item Opaque handles such as \texttt{MPI\_Comm}.
          Implementations can define these to be anything that satisfies the required properties.
    \item Callback functions, e.g., \texttt{MPI\_User\_function}. 
          These callback functions usually do not allow registering any data with the function pointers, 
          which is a challenge to intercepting and forwarding registered functions.
    \item Values for both integer and handle constants, as well as predefined callbacks.  
          Some of these are arbitrary, while others must be chosen carefully.
\end{enumerate}

MPI 4.0 requires that most constants be usable in C for initialization and assignments,
but not case statements, which means they need not be compile-time constants.
Fortran requires they be compile-time constants, which constrains the C ABI
when constants are the same in both languages. 
Buffer address constants cannot be used for initialization/assignment, while
string length constants must be suitable as sizes in array declarations.

\say{ES}{Note: The expression ``not necessarily compile-time constants'' is misleading in the Fortran case because only compile-time constants can actually be used as initialization expressions.}

\say{JH}{The MPI specification discusses this quite a bit.  I do not intend to try to repeat that in the paper, particularly since Fortran isn't the focus of the ABI effort.}

\say{ES}{Note: In practice, many of these constants are implemented as compile-time constants in C, and application codes currently depend on this. For example, codes often use e.g. \texttt{MPI\_THREAD\_SINGLE} as \texttt{case} labels. Of course, this can be remedied straightforwardly.}

\say{JH}{Pre-defined handles are not required to be compile-time constants and Open-MPI does not implement them as such.  I think thread levels are required to be compile-time constants, but if they are not, that will be part of this proposal, as shown below.}

\say{JH}{I created https://github.com/mpi-forum/mpi-issues/issues/642.}

MPICH~\cite{mpich} has elected to provide compile-time constants, 
which is necessary on some operating systems that do not support link-time constants, 
and works in both C and Fortran.
Open-MPI~\cite{gabriel04:_open_mpi} does not have compile-time
constant predefined handles in C, and has an indirection table from
Fortran integer handles to the C ones.

\subsection{MPI integer types}

The types \texttt{MPI\_Aint} and \texttt{MPI\_Offset} are used to store
addresses and file offsets, respectively.
\texttt{MPI\_Count} was added in MPI-3 for the large-count effort, and this
type is required to hold values of \texttt{MPI\_Aint} and \texttt{MPI\_Offset},
so it is at least as large as these.
\texttt{MPI\_Aint} is somewhat challenging since it must hold both
absolute addresses and relative displacements of pointers, so it is similar
to \texttt{(u)intptr\_t} and \texttt{ptrdiff\_t} from C.
However, because it must also work in Fortran as \texttt{INTEGER(KIND=MPI\_ADDRESS\_KIND)},
it must be treated as if it is signed (because Fortran does not support unsigned integers).
Another complication is that pointers, addresses, and differences of pointers may not
always be the same size.  In the past, segmented addressing meant that addresses
could be larger than pointers, whereas there are now platforms where the 
reverse is true, and \texttt{MPI\_Aint} must be able to hold a pointer~\cite{ticket709}
to support struct datatypes, for example.

\subsection{The status object}

This section describes multiple implementations of the \texttt{MPI\_Status} object
and their history.

\subsubsection{New MPICH (MPICH ABI Initiative)}

Below is the status object in MPICH, which was made consistent with Intel MPI, 
in order to establish the MPICH ABI initiative. 
This meant that applications and libraries compiled against Intel MPI 
could be run using many implementations.

\begin{lstlisting}{language=c}
typedef struct MPI_Status {
    int count_lo;
    int count_hi_and_cancelled;
    int MPI_SOURCE;
    int MPI_TAG;
    int MPI_ERROR;
} MPI_Status;
\end{lstlisting}

\subsubsection{Old MPICH}

Prior to being consistent with Intel MPI, MPICH had the following status object.
This definition included unused fields as a hedge against future needs, but
also allowed for platform-specific fields, which meant that MPICH builds on
different platforms could be ABI-incompatible.

\begin{lstlisting}{language=c}
...
typedef struct MPI_Status {
    int MPI_SOURCE;
    int MPI_TAG;
    int MPI_ERROR;
    MPI_Count count;
    int cancelled;
    int abi_slush_fund[2];
    @EXTRA_STATUS_DECL@
} MPI_Status;
\end{lstlisting}

\subsubsection{Open~MPI}

The status object from recent versions of Open~MPI is shown below.
The status used by Wi4MPI has the same layout.

\begin{lstlisting}{language=c}
typedef struct ompi_status_public_t MPI_Status;
struct ompi_status_public_t {
    int MPI_SOURCE;
    int MPI_TAG;
    int MPI_ERROR;
    int _cancelled;
    size_t _ucount;
};
\end{lstlisting}


\subsubsection{MPItrampoline}

MPItrampoline defines a status object that holds the three public fields as well as a union of structs equivalent to the status objects of MPICH and Open MPI.


This definition is not space efficient but convenient for converting between the trampoline
definition and the underlying implementation one, although it stores the public fields
redundantly.

We see here that all variants have the required fields, 
\texttt{MPI\_SOURCE}, \texttt{MPI\_TAG} and \texttt{MPI\_ERROR}, 
and the old MPICH ABI matched the Open~MPI ABI in having both at least one bit for 
the canceled state plus a count field that supports at least 63\, bit values.
The question for ABI standardization is what sort of hidden fields may need to exist
in the future, since there is little to no slack space to add new fields in the current implementations.

\subsection{MPI handle types}

MPI datatypes are opaque objects although the constraints on them limit the implementation choices.
The MPI standard requires that opaque objects can be compared for equality and inequality. 
For the C language, this means that they need to have a built-in type, which reasonably only 
allows integer and pointer types, and excludes union and struct types.

The other important constraint on handles is related to attributes:
\emph{``Attributes in C are of type \texttt{void*} [\dots]
Attributes are scalar values, equal in size to, or larger than 
a C-language pointer. Attributes can always hold an MPI handle.''}
Because MPI handles must be able to be held in a type \texttt{void*},
they cannot be larger than a pointer.

Since Fortran only supports signed integers, and older versions of C provide a limited
set of integer types, one can expect implementations to use a 32-bit integer,
a 64-bit integer, or a pointer for handles, 
although an 8- or 16-bit integer would be permitted.
We see that MPICH uses a C \texttt{int} (32-bits on all supported platforms)
and Open~MPI uses incomplete \texttt{struct} pointers.
The utility of incomplete \texttt{struct} pointers is that they
allow for compiler type-checking.  That is, \texttt{MPI\_Comm}
and \texttt{MPI\_Group}, for example, are recognizable as different types
and the compiler can issue warnings about invalid handle arguments.
On the other hand, the MPICH design allows for zero-overhead conversion
between C and Fortran, as well as the encoding of information in the
handle values themselves. 
Open~MPI does not utilize this capability 
since handles to C objects are not compile-time constants.

Below are some of the MPICH datatype handles, which reveal 
how information is encoded within them:
\begin{lstlisting}{language=c}
typedef int MPI_Datatype;                                                                
#define MPI_CHAR           ((MPI_Datatype)0x4c000101)
#define MPI_SHORT          ((MPI_Datatype)0x4c000203)
#define MPI_INT            ((MPI_Datatype)0x4c000405)
#define MPI_LONG           ((MPI_Datatype)0x4c000807)
#define MPI_FLOAT          ((MPI_Datatype)0x4c00040a)
#define MPI_DOUBLE         ((MPI_Datatype)0x4c00080b)
\end{lstlisting}
These handles encode the size of built-in datatypes that
can be queried trivially with this macro:
\begin{lstlisting}{language=c}
#define MPIR_Datatype_get_basic_size(a) (((a)&0x0000ff00)>>8)    
\end{lstlisting}
There are other macros that take advantage of the hidden structure of 
the \texttt{MPI\_Datatype} handle that the reader can study in \texttt{mpir\_datatype.h}.

Open~MPI's \texttt{mpi.h} defines the datatype handle to be a pointer to an  incomplete \texttt{struct}, which is resolved externally at link-time. The definition of the structure is only visible when building the MPI library itself; otherwise, the compiler only knows its name. This means that the data pointed to by a handle need not be the same at runtime, because the MPI application or library does not depend on it.

\begin{lstlisting}{language=c}
#define OMPI_PREDEFINED_GLOBAL(type, global) ((type) ((void *) &(global)))
...
typedef struct ompi_datatype_t *MPI_Datatype;
...          
#define MPI_CHAR OMPI_PREDEFINED_GLOBAL(MPI_Datatype, ompi_mpi_char)                  
#define MPI_DOUBLE OMPI_PREDEFINED_GLOBAL(MPI_Datatype, ompi_mpi_double) 
...
extern struct ompi_predefined_datatype_t ompi_mpi_char;
extern struct ompi_predefined_datatype_t ompi_mpi_double;
\end{lstlisting}

The runtime cost of querying handles is different in Open~MPI
relative to MPICH.  Open~MPI has to look up the size of the datatype 
inside of a 352-byte struct, which is not a concerning overhead since 
the type of MPI code that will notice such an overhead is going to pass 
the same datatype over and over, in which case the CPU is going to cache 
and correctly branch-predict the lookup and associated use every time.

\begin{lstlisting}{language=c}
static inline int32_t 
opal_datatype_type_size(const opal_datatype_t *pData, size_t *size) {
    *size = pData->size;
    return 0;
}
\end{lstlisting}


Wi4MPI defines all the opaque handles to be \texttt{size\_t}.
This ensures they are at least as large as MPICH's \texttt{int} handles and 
Open~MPI's pointer handles on most platforms
(technically, \texttt{intptr\_t} must be used for this to be strictly true
but the exceptions are obscure~\cite{UCAM-CL-TR-947}).


Wi4MPI defines the built-in datatypes to be sequential integers, 
which means they are not attempting to encode useful information the way MPICH do, 
although they are compile-time constants, unlike Open~MPI. 

\begin{lstlisting}{language=c}
/* C datatypes */
#define MPI_DATATYPE_NULL 0
#define MPI_BYTE 1
#define MPI_PACKED 2
#define MPI_CHAR 3
#define MPI_SHORT 4
#define MPI_INT 5
#define MPI_LONG 6
#define MPI_FLOAT 7
#define MPI_DOUBLE 8
\end{lstlisting}


MPItrampoline uses \texttt{uintptr\_t} internally in its ABI, and 
incomplete \texttt{struct} pointers in its public API for type safety:
\begin{lstlisting}{language=c}
typedef struct MPItrampoline_Comm *MPI_Comm;
typedef struct MPItrampoline_Datatype *MPI_Datatype;
\end{lstlisting}

\paragraph{Analysis}

There are advantages to both approaches.  MPICH optimizes for the common case of built-in types, and does a lookup for others,  while Open~MPI always performs a pointer lookup, but then has what it needs in both cases.

The other advantage of the MPICH approach is with Fortran.  In Fortran, handles are \texttt{INTEGER}
or a type with a single member that is an \texttt{INTEGER}.  
MPICH conversions between C and Fortran are trivial. 
Open~MPI has to maintain a lookup table to map Fortran handles to C objects.

An advantage of the Open~MPI approach of using pointer types to represent opaque types is increased type safety. 
This enables the compiler to flag type mismatches, e.g. an \texttt{MPI\_Comm} 
and an \texttt{MPI\_Datatype} argument have accidentally been swapped. 


\say{ES}{Really? Which should/would/could be deprecated? As far as I can see, the majority of (modern, well-maintained) HPC packages that use the Fortran MPI API still use the old Fortran 77 interface.}

\say{JH}{If they use mpif.h, they are not modern.  That method is not compliant with any standard dialect of Fortran and is just trash on so many levels.  I don't care if it's popular.   It is the leaded gasoline of MPI Fortran.}

\say{ES}{ PETSc.}

\say{JH}{  Then I will fix PETSc.  Also, PETSc is not representative of good or modern Fortran in the slightest.}

\say{ES}{ No, it isn't. But it's representative of how many (supported) packages in HPC look like.}

\say{JH}{  In any case, this paper is about standardizing a C ABI ;-)}

\say{JB}{  yes fortran would lead to so many issues (.mod, multiple versions, multiple compilers, mangling). We only need the C ABI to support portable language bindings. And fortran could become one (external) in fact if the C is stabilized (open question, not sure of all implications).}

\say{JB}{ also ABI is not enough, we need to standardize library names (at least having the main libmpi.so). This is important for containers to be able to inject the host MPI (schematically through repetitive ldds). Not sure of how MPI is drop-in replaceable in a given application through \texttt{LD\_LIBRARY\_PATH} change as the way mpicc's magic works the final binary could end-up to be linked to several other ``support'' libraries (a quick look at MPICH and Open-MPI shows libmpi.so.VERS for both a good start).}

\say{JH}{ Yes, of course, we will have to standardize \texttt{libmpi.so.\$VERS}.  This is probably the easiest part of the ABI effort.}

\say{ES}{ We should also standardize the pkgconfig name for MPI, and we should ensure that \texttt{cmake} and other mainstream tools can recognize the MPI implementation in an unambiguous manner.}

\say{JH}{ I refuse to consider anything related to CMake, and pkgconfig seems beyond the scope of what the standard should address.  Any reasonable build system should be able to figure things out from \texttt{libmpi.so.\$VERS}.}

\say{JB}{ the reason for the FindMPI thing in cmake for example is the mpicc wrapper where you need to get flags from (mpicc -show sort of). The reason for mpicc is the lack of standard library naming no ? if the library is known as most other packages is mpicc usefull anymore ? I would not mention a build-system but maybe the package name should be "mpi" for example. }

\say{GB}{ My preference is to try to restrict ourselves to the minimum viable product which is the ABI. Things like .so name, pkgconfig, cmake, etc. are things that we could consider in the future, but unless they impact the ABI I'd rather move those to an "Outlook" section. }

\subsection{Functions}

Function prototypes in MPI follow naturally from the definitions of their arguments,
which are either opaque handles, MPI integer types, or intrinsic language types.
What is essential for ABI purposes is that the calling convention be fixed.
This can be done by specifying the aforementioned types and defining the
calling convention to be ``as if'' compiled by the platform C compiler.
In most cases, all of the C compilers on a given platform share a calling 
convention but there are at least historical cases where this was not true.
As long as the MPI library uses the platform C compiler calling convention,
it will be compatible with libraries and applications built with it, or
another compatible compiler.

\say{GB}{MPI function declarations follow the symbols and call ABI "as if" their corresponding definition would have been compiled for C. On different platforms or targets this might lead to different symbol convention or call ABIs but within the same target they are the same (TODO: Add example of this). This means that symbol versioning becomes a compatible extension over the standard.}

\say{GB}{Unrelated (from long discussion with JeffH): one thing we could do is add a mpi\_abi.h header file to the standard containing declarations that follow the stable ABI. These would use opaque incomplete types that provide typesafety but fail to compile when dereferenced or introspected. We can define the names of these types in that header. This allows implementations to keep their current ABI in the mpi.h header, but requires them to provide the stable ABI in the separate header if they want to be conforming. This moves the current burden from all users that want to write a portable MPI wrapper to support all implementations, to once per implementation if the implementation wants to be conforming. This allows implementations to innovate internally and decide how much they want to pay for providing a thinner or fatter shim to the stable ABI, and to transition internally from their old ABIs to the new one at their own pace depending on user needs.}
\section{Ecosystem Impact}

One of the main motivations for an ABI is the ability to simplify the end user's life, thus improving the usability of the various MPI implementation through standardization. In this section, we detail particular points of interest for the community which would directly benefit from the availability of an ABI.


\subsection{Python}

The Python language provides MPI bindings through the mpi4py package \citep{mpi4py}. mpi4py uses Cython \citep{cython}, a super-set of the Python language with C extensions. The Cython compiler generates C code calling into the Python C-API and the MPI C-API. The wrapper C code has to be compiled and linked against a specific MPI implementation to generate a Python extension module. 

The lack of a standardized MPI ABI presents several drawbacks. The mpi4py testing infrastructure built on publicly available services like GitHub Actions and Azure Pipelines requires adding both MPICH and Open MPI to the build matrix, effectively duplicating the required resources for running continuous integration. The mpi4py maintainers cannot distribute pre-built binary Python wheels via the Python Package Index, effectively forcing end users to set up a working C and Python development environment and build mpi4py from a sources distribution. The conda-forge \citep{conda-forge} project somewhat alleviates these issues by featuring the conda package manager and its ability to install different variants of pre-built binaries in user-defined non-system locations. Nonetheless, the lack of a standardized MPI ABI prevents {conda-forge} binaries from using MPI implementations that are not ABI-compatible with either MPICH or Open MPI. In addition, conda-forge also suffers from  the doubling of required resources to generate binaries for every downstream application or library using MPI.

A standardized MPI ABI would allow mpi4py to explore alternative implementations based on the runtime loading of dynamic/shared libraries and C foreign function interface (FFI) mechanisms. Such an approach would circumvent the generation of platform-specific binaries, allowing any pure Python code to access the MPI library and its features in a platform-agnostic way.

\subsection{Julia}

The Julia language provides MPI bindings through the \texttt{MPI.jl} package \citep{MPI.jl}.
Unlike Python, Julia does not make use of a C compiler to call into external libraries.
Instead, the user provides the corresponding types to the function signature to the \texttt{ccall} command.
To support this, the developers of \texttt{MPI.jl} had to (a) define the constants and type definitions for each MPI ABI, (b) develop heuristics to detect which ABI a particular MPI library is using, and (c) provide a mechanism to switch between the ABI definitions, invalidating Julia's cache of pre-compiled code. 
This code has been a significant source of issues, hampering its usability and requiring significant engineering effort on the part of its volunteer maintainers.

A key benefit of a standardized ABI will be making it easier to provide downstream binaries. The Julia package manager provides prebuilt binaries of many MPI-enabled libraries, such as ADIOS2, HYPRE, P4est, and PETSc,
but the support is rather cumbersome, especially when users wish to use non-bundled MPI implementations. The ABI would boost usability, especially for the long-tail of users on lower-end systems.

\subsection{Rust}
\tofix{Rust (Jed or Gonzalo?)}
\say{GB: I've given this a pass. I think we may want to add that this would make it easier to use libraries that depend on MPI from rust, like HDF5, etc.}

The Rust programming language provides MPI bindings through the libraries in the \texttt{rsmpi} project \cite{rsmpi-github}.
\texttt{rsmpi} combines a thin static library that re-exports underspecified identifiers (providing symbols where MPI implementations are allowed to use macros) and uses \texttt{bindgen} to create the raw Rust interface.
\texttt{bindgen} relies on \texttt{libclang} to parse the header (of the thin library and \texttt{mpi.h}) to generate Rust bindings that conform to the C ABI of the MPI implementation at hand.
This approach works for any compliant MPI and does not require tedious definition or maintenance of stubs.
The disadvantages are many and include long initial compilation times due to having to fetch and build the dozens of dependencies of \texttt{bindgen}, a need for pre-installed \texttt{libclang}, long testing times by building against multiple different MPI implementations, and need for users to rebuild to pick a different MPI implementation.

Rust is known for reliable package management and tooling for binary distribution, including cross-compilation (across OS and ISA).
A standardized MPI ABI would allow \texttt{rsmpi} to provide a thin and stable Rust binding that can be built without any dependencies and simply links against a dynamic library on the target platform.
For example, a single CI/CD job for an application could publish binaries for MacOS (x86-64 and ARM), Android, Windows, and Linux (x86-64, POWER, and ARM) without needing to think about MPI idiosyncrasies.
ABI stability would make it easier to support more MPI features and would also enable low-level idiomatic Rust features that improve safety and static analysis for C FFI bindings, but that are hard to incorporate into the current \texttt{bindgen} approach.

\subsection{Fortran}

Currently, all implementations of MPI Fortran wrappers are integrated with MPI implementations.
Vapaa~\cite{vapaa-github} is the first attempt to write the MPI Fortran 2008 interface
as a standalone project, based on calling the C interface, without any use of the internal state.
Because Fortran constants must be compile-time constants, not just link-time constants,
when Open~MPI is used, Fortran interfaces must define their own set of constants and translate
them to the C ones at runtime.
Furthermore, to handle status objects, the status ABI must be known.
Thus, Vapaa ends up implementing its own ABI and translating all constants and status objects.
A standard ABI would simplify the translation process and eliminate the need for implementation-specific status handling.
If the Fortran 2008 interface had the type of \texttt{MPI\_VAL} equivalent to handle types
in C, then no translation would be necessary.
However, this would be both an ABI and an API change for \texttt{mpi\_f08.mod};
it also offers nothing to users of older MPI Fortran interfaces.

In addition, due to differences in name mangling among Fortran compilers, running a Fortran program that calls MPI functions inside a container can result in linker dependencies that are not fixed and depend on the specific Fortran module and compiler convention used. This prevents an MPI Fortran program from having its MPI implementation replaced through interposition (i.e., \texttt{LD\_PRELOAD}). Having an external Fortran implementation that relies on the ABI would enable the static linking of the Fortran adaptation layer in the target binary, abstracting away from these language-dependent variations and restoring ABI compatibility. 

\subsection{Packaging}

The availability of an ABI is highly important for packaging MPI applications. MPI is a fundamental package for most scientific software. However, there are several libraries that provide MPI support, including vendor-specific ones. Therefore, building a binary for MPI can become cumbersome when managing a long chain of dependencies between packages, resulting in repetitive building. While the ABI alone is insufficient to solve packaging dependency issues, it is a significant step in the right direction. When running an executable, the loader is responsible for locating various dynamic shared objects (DSOs) to fulfill execution dependencies, based on the system's configuration and environment (e.g., \texttt{LD\_LIBRARY\_PATH} and search paths). Thus, the ABI alone has no impact on the loading of libraries when running a program. Nonetheless, it is still a crucial step towards achieving drop-in replacement for MPI, which involves changing the MPI of a given binary. This goal can be attained by defining a common library naming scheme or developing specific stub libraries in charge of bridging implementations -- the binary being linked to the stubs.

\tofix{Spack (Todd)}

\tofix{easybuild}

\tofix{Linux distros (Ubuntu and Red Hat)}

Linux package managers such as APT and RPM ship binaries for two dependency chains, with packages like \texttt{hdf5-openmpi} and \texttt{hdf5-mpich}.
APT manages these through the \texttt{/etc/alternatives} mechanism while RPM (Fedora) deliberately rejects that usage due to their rule \cite{fedora-alternatives}: ``If a non-root user would gain value by switching between the variants then alternatives MUST NOT be used.''
As such, Debian/Ubuntu users developing code that depends on an MPI-enabled HDF5 can have a default implementation (that might change unexpectedly at the whims of their sysadmin) while Fedora/Red Hat users must use verbose paths unlikely to be found by configure scripts.
In this example, Arch Linux provides \texttt{hdf5-openmpi} in the default repository (binary distribution) while \texttt{hdf5-mpich} (and \texttt{mpich} itself) are in the user repository that must be installed from source.
Homebrew provides only \texttt{hdf5-mpi}, which uses Open MPI.
This complexity requires maintenance and communication to users, increases testing time and frequency of bugs, and harms reproducibility.
With an ABI, there could be one \texttt{hdf5-mpi} and let the \texttt{mpiexec} or ldconfig/\texttt{LD\_LIBRARY\_PATH} (since these distributions eschew \texttt{RPATH}) determine which MPI to execute with.

\tofix{NVHPC (Jeff's people)}

\tofix{conda-forge (Lisandro)}

A possibility to retarget binaries would be changing embedded \texttt{RPATH}s inside the executable. Spack~\cite{gamblin2015spack}, which relies extensively on \texttt{RPATH}s has implemented such rewriting techniques~\cite{zakaria2022mapping} when deploying binary packages to systems, relocating the \texttt{RPATH}s through carefully planned compilation and clever rewriting techniques. This approach was required as the end-user may have deployed his spack tree at a different location from where it was initially compiled. By doing so, Spack then manages to restructure complex dependency chains, it is a process analogous to what would be required to change the runtime of a given MPI-enabled binary.
The Anaconda \citep{anaconda} software distribution and its conda package manager also rely on \texttt{RPATH} rewriting to allow binary relocation.

\subsection{Testing}

\tofix{mpi4py (Lisandro)}

The advantage of the ABI support for testing is less direct. Indeed, when running a program with a given MPI, the MPI is also part of the equation to be validated. Gains could be envisioned in the building of the test cases but it is not obvious that MPI implementation will be directly interoperable even in the event of a unified ABI. Indeed, as for packaging, the build chain for MPI involves compiler wrappers and implementations are free to choose the name of the MPI Dynamic Shared Object (DSO) -- preventing drop-in replacement of MPI. Overall, for testing, the ability to retarget MPI programs to another implementation requires more than an ABI but either a clearly defined object layout for MPI or a dedicated redirection layer similar to what is provided in ``trampoline'' interfaces~\cite{Wi4MPI-paper,MPItrampoline}. Analogous dependencies on the naming of the DSO are also present for containers as discussed in the next section.

\tofix{PETSc (Jed)}

\subsection{Containers}

Containers are an abstraction built on Linux namespaces.
Containers are the systematic use of such namespaces to run ``software images'' in a custom environment.
The main advantage of containers is the ability to move images around systems to avoid recreating complex software environments. 
In HPC networks, it is common to use OS bypass techniques to optimize network performance. This involves creating a separate
networking layer that operates independently of the operating system, allowing for faster and more efficient communication between nodes. As a result, the networking namespace (IP bound) is not commonly used in HPC networks which prefer faster fabrics than TCP/IP. Similarly, to prevent security issues such as privilege escalation, the user namespace, which allows mimicking root behavior inside the container, is not used for containers in non-virtualized environments\cite{pcocc}. Overall, the namespace leveraged by HPC container runtimes such as Singularity and Shifter is the mount namespace.

Using the mount namespace, it is possible to change the mount point seen by the running program. 
In HPC, the user's home is often bind-mounted inside the container, stacking the container's view of the file system atop of the preexisting shared one. 
With a containerized program compiled against MPI, the corresponding MPI is likely present in the container image. 
This MPI has  to be compatible with a wide range of interconnects, unlike the host MPI from the system, since it cannot anticipate its target environment. 
While this can be mitigated with communication libraries such as libfabric\cite{libfabric}, which manage to unify high-speed network interfaces, there may be features that are not possible outside of the native MPI environment, perhaps because they are proprietary and not generally available, or because they require system awareness (e.g., network topology information) that cannot be included in the widely distributed implementations of MPI.

On this aspect, using the host MPI (as opposed to a container MPI) 
would allow the guest binary to take advantage of all the custom features of the system. 
It also obviates the need for application containers to redistribute MPI at all.
For this purpose, approaches such as \texttt{e4s-cl}~\cite{skutnik2021e4s} 
recursively locate all dependencies of MPI and inject them into the container.
The target binary may depend indirectly on libraries such as \texttt{hwloc} that are required by MPI.
There are ways to mitigate this issue, like embedding all dependencies or using symbol versioning for standard HPC libraries to reduce possibilities for symbol conflicts. 
Another method is to ensure that MPI-shared libraries do not cause transitive dependencies
so that the binary only requires MPI, and the MPI implementation takes care of its dependencies directly.
A second challenge faced by the MPI container is launching the application. 
Indeed, MPI is in many cases relying on the Process Management Interface (PMI) to wire up its processes, 
which also has to be mapped into the container. 
There have been studies on this point as part of a complete rework of this interface in the PMIx standard~\cite{castain2017pmix} enabling PMI portability.

To summarize, containers need support from MPI to allow binary retargeting, 
i.e., the ability to change the MPI implementation on a binary compiled against another MPI implementation. 
Note that changing the guest MPI to the host MPI also allows PMI disambiguation 
-- removing complexity on the launch side. 
Having an ABI is compulsory as retargeting does not allow recompilation of the application. 
This last point is the main blocker for accepting and distributing MPI containers.

\say{MP}{I'm not sure, but standardized MPI ABI is insufficient for injecting MPI in an existing binary in a container. We have to be sure that other libraries (for example, HWLOC) used by the MPI library and the application also have standardized ABI. Am I wrong? If not, we can mitigate the proposed text. (Sorry for this late remark)}

\subsection{Performance and Debugging Tools}

MPI tools often use the profiling interface (PMPI) to intercept function calls
and extract the current state of MPI and to time operations, for
performance and debugging purposes~\cite{knupfer2012scorep, hilbrich2010must} .
Since this interception operates on the compiled library code,
all MPI tools must be compiled against the relevant implementation ABIs.
A standard ABI makes it possible for PMPI interposition tools
to be compiled only once and reused with different MPI implementations.

The Tools working group in the MPI Forum is working on the QMPI interface~\cite{elis2019qmpi}. 
This interface is designed to support multi-instrumentation, 
mimicking what has been previously pioneered with $P^{n}MPI$~\cite{schulz2008pnmpi}. 
As with PMPI, the absence of a standard ABI requires each QMPI tool to be compiled
for every ABI, and potentially more, if any of these tools modify 
ABI-related properties of the interface.
One of the advantages of the proposed status object for the standard ABI
is that it has additional space that allows tools to hide state
in the reserved fields.
Managing this in a layered context is complicated and is left as an exercise
for the implementers of such tools.
\section{Proposal}

This section outlines the current proposal for the MPI standard ABI,
based upon detailed analysis of requirements from MPI as well as
the behavior of platforms MPI should support.
%
%
Following Section~\ref{sec:current-abi}, the ABI proposal defines
MPI integer types,
the status object,
opaque handles,
and constant values.
The calling convention must be equivalent to the platform C compiler.

\subsection{MPI integer types}

The purpose of \texttt{MPI\_Aint} is to hold addresses or pointers,
whichever is larger, because its usage requires both.
It should also be signed because Fortran does not support unsigned integers.
The only standard C type that meets this requirement is \texttt{intptr\_t}.
It is necessary to use this integer type on platforms with so-called 
``wide pointers'' \cite{UCAM-CL-TR-947}, although this situation is rare.
There is no C integer type associated with filesystem offsets, but all
modern systems should use at least 64-bit integers.
There are some platforms where the underlying filesystem offset
may be 128-bits, but there is no need for MPI to define \texttt{MPI\_Offset}
this way since MPI files greater than 8 EiB are unlikely.\footnote{
    At current prices of \$10/TB, one such file would require more than \$90M in filesystem hardware.
}
Additionally, 128-bit integers are not implemented natively on most systems 
and thus may perform poorly, so it is undesirable to force the use of
128-bit integers for offset and count to support impossibly large files.
On the other hand, most systems with 32-bit addressing have
64-bit filesystems, so there are at least some scenarios where the MPI ABI
should be flexible enough to support different sizes of address and offset types.

To ensure all relevant target platforms can be supported,
the MPI ABI should be described in terms of the size of
\texttt{MPI\_Aint} and \texttt{MPI\_Offset},
while \texttt{MPI\_Count} matches the larger of these two
(which will be \texttt{MPI\_Offset} on most systems).
The integer sizes of the MPI ABI can be denoted
A$n_{\rm Address}$O$n_{\rm Offset}$, to denote the number of bits
in the \texttt{MPI\_Aint} and \texttt{MPI\_Offset} types, respectively.
This is similar to how platform ABIs are described using
I$n_i$L$n_l$LL$n_{ll}$P$n_p$ notation, to denote the size of
C \texttt{int}, \texttt{long}, \texttt{long long}, and
\texttt{void*}, respectively.
For example, modern Linux platforms are described as LP64, 
meaning that \texttt{long} and \texttt{void*} are 64-bit.
Today, essentially all MPI ABIs are A32O64 or A64O64 ABIs,
because we have 32- or 64-bit addresses, but most filesystems are 64-bit.
An A64O128 ABI is possible, although, for the aforementioned
reasons, it is neither necessary nor desirable.

The potential for more than one MPI ABI on a given platform is undesirable.
Current trends in filesystem technology suggest that a \texttt{MPI\_Offset} 
larger than 64 bits will not be necessary for at least 20 years.
For these reasons, we propose to prescribe the MPI ABI for platforms 
with 32- and 64-bit pointers as follows:
\begin{lstlisting}{language=c}
typedef intptr_t MPI_Aint;
typedef int64_t MPI_Offset;
typedef int64_t MPI_Count;
\end{lstlisting}
This ABI definition covers essentially all relevant platforms
since the introduction of LFS~\cite{LFS-wikipedia}
until the availability of filesystems far in excess of 8 exabibytes.
These types are part of C99 and C++11 but implementations
can use older equivalents for compiler portability.

One may observe that \texttt{intptr\_t} is optional in C and, in theory,
a system may lack an integer type capable of satisfying its requirements.
This is uncommon, and exists to accommodate systems
with 128-bit pointers but where supporting \texttt{intptr\_t} would
force a change in \texttt{intmax\_t}, which would be a breaking change
in the platform ABI~\cite{n2873}.
We note that MPICH requires \texttt{intptr\_t} and platforms that
do not provide it are not supported, so a reasonable portion of
the MPI ecosystem is unconcerned with this situation.

While we have considered the case of 128-bit pointers,
the current proposal will only include A32O64 and A64O64.
It is appropriate for the MPI community to gain more experience
with such platforms before attempting to standardize for them.
For example, while CHERI~\cite{UCAM-CL-TR-941} has 128-bit pointers
but doesn't necessarily require 128-bit file offsets, but if
\texttt{MPI\_Aint} and therefore \texttt{MPI\_Count} have to be
128b, it might be prudent to make offsets the same width, so that
there is only one MPI ABI for all 128-bit platforms.

The one MPI integer that cannot be prescribed like the others is
\texttt{MPI\_Fint}, since this corresponds to Fortran \texttt{INTEGER},
which is not fixed, but varies as a function of Fortran compiler flags.
It seems appropriate to have a runtime query to allow C code to know 
the size of a Fortran integer and work with it appropriately.
This requires code changes compared to the current situation where
\texttt{MPI\_Fint} is known at compile-time, but the C code that
relies on this is rare.
Alternatively, the standard ABI could force \texttt{MPI\_Fint} to
be a C \texttt{int}, and disallow MPI Fortran interfaces from supporting
larger integer sizes.
This would please Fortran purists who loathe the compiler feature that
allows changing the Fortran default integer size, but displease
users of existing implementations that support it.

\subsection{The status object}

The proposed standard status object is:
\begin{lstlisting}{language=c}
typedef struct MPI_Status {
    int MPI_SOURCE;
    int MPI_TAG;
    int MPI_ERROR;
    int mpi_reserved[5];
} MPI_Status;
\end{lstlisting}

This object is 32 bytes in size, which leads to good alignment when arrays
of statuses are used, and includes at least two extra fields more than 
current implementations.

\subsection{Handle types}

In order to have type-safety in handles, incomplete struct pointers are proposed; 
Open~MPI has used this design and its properties are well understood.
The incomplete struct name will become part of the ABI, so that compiler warning messages are clear:
\begin{lstlisting}{language=c}
typedef struct MPI_ABI_Comm * MPI_Comm;
typedef struct MPI_ABI_Request * MPI_Request;
\end{lstlisting}

\subsection{Constants}

Constants in MPI come in different forms.  They include:
\begin{itemize}
    \item Error codes, which start with \texttt{MPI\_SUCCESS=0}.
    \item Buffer address constants, e.g., \texttt{MPI\_BOTTOM}, 
          which must have special values distinguishable from user buffers.
    \item Handle constants.
    \item Integer constants that must have special values to avoid conflicts;
          for example, \texttt{MPI\_ANY\_SOURCE} can never be a valid rank,
          and thus should be a negative number.
    \item Integer constants that must be powers of two, to support combination using XOR.
    \item Integer constants that correspond to string lengths.
    \item Integer constants that can have any value.
    \item Predefined attribute callback functions.
\end{itemize}

Some of the desirable properties brought forth by users and implementers include
a desire for unique integer constants, so that errors can be identified precisely.
For example, if a user passes \texttt{MPI\_ANY\_TAG} as a rank, this can be identified
precisely if the constant value is unique with respect to all other constants, especially
\texttt{MPI\_ANY\_SOURCE}.
Another desirable property is the ability to encode information in handle constants,
as MPICH does.
For maximum portability, integer constants cannot be larger than 32767, because
that is the largest value of type \texttt{int} guaranteed by the C standard.
This constraint is strictly academic for the relevant systems but there was
no reason to violate it either.

For handle constants, the working group discussed designs with and without
unique values as well as the use of one or more lookup tables versus a Huffman code.
The current proposal uses a Huffman code but is sufficiently compact so
as to require a relatively small lookup table, for implementations that choose to use one.
The Huffman code uses 10 bits and therefore fits into the zero page of common
operating systems; as a result, implementations that allocate user handles
from the heap need not verify that they do not conflict with predefined constants.

As datatypes make up the majority of MPI's predefined handles, half of
the Huffman code bits are reserved for datatypes, although less than 100
values are used.
The language, numerical properties, and sizes of all \textit{fixed-size}
types are encoded in the handles.
For example, \texttt{MPI\_CHAR} can be determined to be a 1-byte
type immediately.
Unfortunately, \texttt{MPI\_INT} is not a fixed-size type, so its size
is not encoded, as that would mean that the constant value was a function
of the platform ABI.
While it would be possible for some use cases to handle this, it is
undesirable to force higher-level languages like Julia to determine
the platform ABI in order to use MPI.

Other handles can be decoded quickly using the bit pattern alone.
The value zero is always an invalid handle, which allows uninitialized
handles to be detected as errors instead of being confused as legal null handles.
Legal null handles use the non-zero bits of the handle kind followed by zeros.
The current Huffman code has a sufficient amount of free space to allow
for many new handle types and new handle constants for existing types
to be added, without requiring breaking changes.

The values of integer constants for string lengths,
e.g., \texttt{MPI\_MAX\_\-LIBRARY\_VERSION\_STRING},
and constants that can be combined with XOR,
e.g., \texttt{MPI\_MODE\_NOCHECK},
are not particularly interesting.
For the former, the largest known values used in
existing implementations were chosen.
There was some concern that stack allocation of 8192 bytes
could be a problem, but 
(1) nothing prevents users from allocating such strings on the heap and 
(2) no issues with this value (used by MPICH) have ever been reported.

The other integer constants are unique negative numbers, which means
that implementation can tell the user by name what constant they passed,
when the user passes an incorrect constant.

For simplicity, predefined attribute callbacks were set to \texttt{0x0} for \texttt{MPI\_XXX\_NULL\_COPY\_FN} and \texttt{MPI\_XXX\_NULL\_DELETE\_FN}, and \texttt{0xD} for \texttt{MPI\_XXX\_DUP\_FN}.
Since compilers can detect incompatible function pointer arguments there is little need to detect errors at runtime.

The encoding of operation handles is provided in Appendix~\ref{app:op_handles}.
The gaps in the ranges for the different operation types are intentional since they provide room for future extensions.
Moreover, the modified Huffman encoding enables fast error checking by implementations, simply by applying a bitmask.

Handles for opaque objects are encoded in a similar way, as shown in Appendix~\ref{app:opaque_handles}.
The encoding leaves room for future extensions for each handle type, making it possible to add new handles without requiring special case handling.

Examples of datatype handles are provided in Appendix~\ref{app:datatype_handles}.
Types with variable size (e.g., C \texttt{int}, \texttt{float}) 
are encoded with the prefix \texttt{\textbf{0b1000}XXXXXX}.
Fixed-size types are encoded with the prefix \texttt{\textbf{0b1001}XXXXXX}, 
with the size encoded in the lower bits at position 4--6.
For example, types with size 1 are encoded with prefix \texttt{0b1001\textbf{000}XXX} 
(e.g., \texttt{MPI\_BYTE} with \texttt{0b1001\textbf{000}111}; 
size $2^{000b}$) while types with size 4 are encoded with prefix \texttt{0b1001\textbf{010}XXX} 
(e.g., \texttt{MPI\_INT32\_T} with \texttt{0b1001\textbf{010}000} and size $2^{010b} = 2^2$).

The full definition of the Huffman code for handle constants can be
found in \cite{print-handle-constants}, while the other constants
are listed in \cite{IntegerConstants}.

\section{Experiments}

In this section, we present three experiments regarding the
implementation of the standard ABI.
First, we measure the performance impact of different ABIs
for querying the size of a type.
Second, we measure the message rate for MPICH-based implementations,
with and without standard ABI support.
Third, we describe Mukautuva, which demonstrates the feasibility of implementing the
standard ABI outside of any existing implementation.
Finally, we mention the effort required to implement the standard ABI in MPICH.
For both implementations -- the one outside of an MPI implementation
and the one within MPICH -- we see that the cost of ABI translation is small.

\subsection{Performance}\label{sec:perf}

Historically, there has been a performance argument in favor of MPICH's
integer handles for datatypes because information like type size is
encoded directly in handles, whereas with Open~MPI, it must be fetched
from the internal state.
We measured the throughput of \texttt{MPI\_Type\_size} to be be $\approx 11.5$ nanoseconds
with both implementations on an AMD EPYC 7413 CPU.
Not only is the difference between the two implementations negligible,
both are negligible relative to the network cost of sending a single message,
which is at least 500 nanoseconds.

Table~\ref{table:osu-mr} shows the message rate determined by the OSU MPI Benchmarks 7.0.1
for three different builds of MPICH: 
the latest Intel MPI
and
MPICH development versions built with UCX
using the MPICH ABI\footnote{
    https://github.com/pmodels/mpich \texttt{0f72280c70214411e00e5c03e1f4111972fab4b7}
}
and the standard ABI prototype\footnote{
    https://github.com/hzhou/mpich/tree/2302\_abi \texttt{4ce7cd3b6c6a92e99090957ae1c4a9166efadedf}
}
, with and without Mukautuva.
We see that adding the indirection from Mukautuva has a noticeable impact,
but it is likely acceptable as a worst-case implementation of the standard ABI.

\begin{table}
\caption{Message rate (8-byte messages) determined by \texttt{osu\_mbw\_mr} 
         on an Intel i7-1165G7 CPU 
         running Linux 5.19.0 (Ubuntu 22.04).
         Build options unrelated to ABI
         -- the shared-memory performance of UCX versus OFI --
         have a significant impact on message rate.
         The MPICH dev UCX results show no difference between 
         the MPICH ABI and the proposed standard ABI.
    \label{table:osu-mr}
}
\begin{tabular}{ |l|r| } 
 \hline
 MPI & Messages/second \\ 
 \hline
 Intel MPI 2021.9.0     &  4658939.64 \\
 + Mukautuva            &  4606473.95 \\
 \hline
 MPICH dev UCX [1]      & 13643117.42 \\
 + Mukautuva            & 12278837.03 \\
 \hline
 MPICH dev UCX ABI [2]  & 13643378.98 \\
 \hline
\end{tabular} \\
1. \texttt{--enable-error-checking=no --enable-fast=Os --enable-g=none --with-device=ch4:ucx} \\ 
2. Same as 1 plus \texttt{--enable-mpi-abi} 
\end{table}


\subsection{Mukautuva}

Mukautuva~\cite{mukautuva-github} (``Adaptable'' in Finnish)
was created both as an ABI compatibility layer
and as a way to prototype the ABI proposal being developed for the MPI Forum.
It represents a worst-case scenario implementation for the standard ABI,
if implementers refuse to support it.

Mukautuva (MUK) consists of two shared libraries.
The first library provides the MPI interface symbols.
The second library is compiled against
MPICH or Open~MPI and provides the underlying implementation (IMPL).
Applications compiled against MUK are relying on its ABI,
which is a proxy for a future MPI standard ABI.
At runtime, the first shared library determines which implementation
will be used, and  activates it via \texttt{dlopen} and \texttt{dlsym}.
MPI symbols call a wrapper layer with the MUK namespace.
MUK symbols are function pointers to the WRAP namespace in the 
implementation-specific shared library.
WRAP functions call the implementation, with the appropriate
conversion of handles and constants.
An excerpt for the case of \texttt{MPI\_Comm\_size} follows.

\begin{lstlisting}{language=c}
libmuk.so:

typedef union {
    void *      p;  // Open-MPI
    int         i;  // MPICH
    intptr_t    ip;
} MUK_Handle;
typedef MUK_Handle MPI_Comm;

// during initialization
...
MUK_Comm_size = MUK_DLSYM(wrap_so_handle,"WRAP_Comm_size"); // wraps dlsym()
...

int MPI_Comm_size(MPI_Comm comm, int * size) {
    return MUK_Comm_size(comm, size);
}

impl-wrap.so:

#include <mpi.h> // implementation details

static inline MPI_Comm CONVERT_MPI_Comm(WRAP_Comm comm) {
    if (comm.ip == (intptr_t)MUK_COMM_WORLD){ return MPI_COMM_WORLD; } else 
    if (comm.ip == (intptr_t)MUK_COMM_SELF) { return MPI_COMM_SELF; } else 
    if (comm.ip == (intptr_t)MUK_COMM_NULL) { return MPI_COMM_NULL; } else 
    {
#ifdef MPICH
        return comm.i;
#elif OPEN_MPI
        return comm.p;
#else
#error NO ABI
#endif
    }
}

// success is the common case, so static inline it.
int ERROR_CODE_IMPL_TO_MUK(int error_c);
static inline int RETURN_CODE_IMPL_TO_MUK(int error_c) {
    if (error_c == 0) return 0;
    return ERROR_CODE_IMPL_TO_MUK(error_c);
}

int WRAP_Comm_size(WRAP_Comm comm, int *size) {
    MPI_Comm impl_comm = CONVERT_MPI_Comm(comm);
    int rc = IMPL_Comm_size(impl_comm, size);
    return RETURN_CODE_IMPL_TO_MUK(rc);
}
\end{lstlisting}

The vast majority of MPI features can be translated from
one ABI to another with trivial overhead.
The exceptions to this come in two forms:
first, when callbacks are involved, and
second, when vectors of handles are required.
For callbacks, MUK must translate to IMPL handles
to call IMPL functions, but then translate IMPL
handles back to MUK handles, because the callback functions
compiled as user code utilize the MUK ABI.
The callback interfaces do not always make this easy,
but it can be done in all cases, using methods described in
the \texttt{README.md}.
The situation with vector arguments is similar to~\cite{bigmpi-paper},
where vectors of datatype handles must be converted from 
one ABI to another, and freed upon completion, which is tricky in the case
of nonblocking \texttt{alltoallw} operations.
For these cases, like with callbacks, we use a map, currently
implemented with \texttt{std::map} from the C++ standard library,
to associate a temporary state with a handle.
Callback function trampolines or request completion operations
lookup the temporary state associated with handles when needed.
The worst-case overhead will arise when the user has initiated
a nonblocking \texttt{alltoallw} operation, followed by a large number
of nonblocking point-to-point operations to be completed
via \texttt{MPI\_Testall}, for example.
In this case, every call to \texttt{MPI\_Testall} will look
up every request in the map associated with nonblocking alltoallw operations.
This is not currently optimized, due to the low probability
of such a scenario in real applications.

During the development of MUK, we identified flaws in the early
ABI proposals as well as in MPI test suites.
The MPICH test suite, for example, assumed the MPICH ABI in 
many places\footnote{
 e.g. https://github.com/pmodels/mpich/issues/6398
}, which meant that it could not be used to test
other implementations, or ABI translation layers such as
Wi4MPI, MPItrampoline, and MUK.
Most if not all of these issues have been resolved in the meantime.

MUK now passes all of the MPICH test suite tests
except for a handful that uses dynamic process management,
which appears to be related to environmental problems, yet to be investigated.
MUK also passes all tests associated with ARMCI-MPI,
the Intel MPI Benchmarks (IMB),
and
the OSU MPI Benchmarks (OMB).
It complies with MPI-4 except for sessions,
which are expected to suffer from the same issues observed
in dynamic process management functionality.
Calling functions before initialization or after finalization
is not fully supported, 
and will be fixed in the future.

\subsection{MPICH}


While it has been demonstrated that the standard ABI can be implemented
without any change to existing implementations, doing the translation
inside of an MPI implementation has lower overheads.
Hui Zhou has implemented support for the standard ABI in MPICH~\cite{mpich-abi-github}.
The changes consist primarily of abstracting away prior assumptions about
the types of handles and callback signatures and inserting the
appropriate conversions, where necessary.
Most of the changes are in the interface code generator or guarded
by a preprocessor token, hence having no impact on execution time.
The most expensive conversions are for datatype and reduce ops,
with a worst-case that requires $O(N_{\textrm{predefined}})$ comparisons.
\section{Other Considerations}

A standard ABI is necessary but insufficient to provide
seamless compatibility of MPI software across implementations.
For example, MPI applications often require a parallel launcher,
e.g., \texttt{mpiexec}, which is not part of the ABI, but 
interacts with the MPI program in non-standard ways,
such as environment variables.

There are at least two solutions for portable launching.
The first is that the launcher determines the MPI
shared library to be used, in which case the launcher
and the library will be compatible.
Another is the use of a launcher that is supported by
multiple MPI implementations, such as the ones provided
by popular schedulers like SLURM and PBS.

Applications also need to know what shared library to use.
As \texttt{libmpi.so} is used by a number of implementations
already, the name \texttt{libmpi\_abi.so} is proposed for
implementations of the standard ABI.
Standardizing a new, descriptive name is especially important
since it is expected that implementations will continue
to support their existing ABIs, using the existing library name(s).
It is expected that \texttt{libmpi\_abi.so}  will follow the 
platform-specific conventions for versioning to allow
for future -- hopefully backwards-compatible -- changes.

Obviously, much of the MPI ABI is contained in the header file, \texttt{mpi.h}. 
The same filename will be used for the standard ABI, to ensure source compatibility, 
but applications must use exactly one ABI, and therefore every component of an 
application will need to be compiled against the same header. 
We expect that the standard ABI will be implemented in a header file provided by 
the MPI Forum that can be used with any implementation that supports the standard ABI, 
to ensure consistency in its definition.
Implementations can provide this header in a different path from their own header,
and perhaps help users with appropriate pkg-config definitions or compiler wrapper scripts,
e.g., \texttt{mpi\_abi.pc} or \texttt{mpicc\_abi}, but these aspects of MPI are not standardized,
nor are they part of the ABI.

\subsection{Fortran}

This paper focused on a standard C ABI for MPI, but many codes use MPI from Fortran.
Fortran presents its own ABI challenges, not the least of which is that
\texttt{INTEGER}, used for MPI handles in \texttt{mpif.h} and \texttt{mpi.mod}
(and the \texttt{MPI\_VAL} in typed handles defined by \texttt{mpi\_f08.mod})
varies in size depending on compiler flags.
Furthermore, each Fortran compiler has its own ABI and each has their own
runtime library, in contrast to C, where it is common for C compilers to
reuse the system C runtime, and thus be ABI compatible (e.g., Intel and GCC on Linux).

The current ABI proposal for Fortran follows the C one;
many constants are required to be the same in both languages anyways.
While Fortran handles may be too small to hold the C handle values in general,
implementations can optimize for the case of predefined handles because
the C constants will be representable in Fortran integers and
do not require a translation table.



The overhead of translation for user-defined handles could be achieved with a
new implementation of \texttt{mpi\_f08.mod}, where \texttt{MPI\_VAL} 
is \texttt{INTEGER(kind=c\_intptr\_t)}, although this is a breaking change and
would require a new module, which could be called \texttt{mpi\_f08\_abi}.
One could also imagine a module \texttt{mpi\_abi} that requires 
handles be \\ \texttt{INTEGER(kind=c\_intptr\_t)}.
No new MPI Fortran interfaces or modules are currently proposed.
\section{Conclusions}

We have reviewed the current practices for MPI ABIs in the popular implementations,
MPICH and Open MPI, as well as ABI abstraction layers like Wi4MPI and MPItrampoline.
The motivations for standardizing an MPI ABI come from multiple sources, including
the packaging and distribution of MPI applications and libraries in binary form,
the use of MPI from languages other than C (or C++),
and the development of implementation-agnostic MPI performance and debugging tools.
The MPI ABI working group has developed a proposal for a standard MPI ABI,
which satisfies all of the requirements and relies only on ISO C language features.
The standard ABI has been prototyped in both MPICH and Mukautuva, and is determined
to be both practical and performant.
We identified issues with compatibility and portability not related to the ABI
that are expected to be solved by the MPI ecosystem.

The next steps for the proposed MPI ABI are
(1) it must be standardized by the MPI Forum,
(2) it must be implemented either by the major implementations and/or
    ABI abstraction layers.
(3) users of MPI must recompile against the standard ABI.
Work towards 1 is underway, and this paper has provided sufficient evidence
that 2 is either already done or straightforward.

We cannot predict the behavior of all MPI users, and certainly, some may be
reluctant, either because they expect the MPI standard ABI to be less reliable
than existing ABIs or that it will break in the near future.
There is obviously a large one-time cost of recompiling everything against 
the MPI ABI, but it is no worse than the cost of compiling everything
against a new major release of Open MPI, for example.
Fortunately, there is no immediate need for users to adopt the MPI ABI.
It is expected that both MPICH and Open MPI will support their existing
ABIs for as long as users require them, and will consider a translation
to using the standard ABI natively only after there is sufficient understanding
of its use across a wide range of platforms.

Regardless of how long it takes to realize the full potential of a standard ABI,
we expect that it will significantly reduce the pain of using MPI in a variety
of contexts, and encourage greater use of MPI in new domains.

\begin{acks}

The authors are indebted to numerous people who
contributed to the MPI ABI standardization effort,
including, but not limited to,
Brian Barrett,
Jim Dinan,
Dan Holmes,
Julien Jaeger,
Quincey Koziol,
Ken Raffenetti,
Jeff Squyres,
and
Bill Williams.
Simon Cooksey was a valuable resource regarding the CHERI architecture,
which exposes some of the sharper edges of the C standard regarding
pointers and integer constants.
Jeff thanks Tim Costa for encouraging him to work on 
ABI standardization in the first place.
Jed's work was partially supported by the U.S. Department of Energy, Office of Science, Office of Advanced Scientific Computing Research, applied mathematics program.
Research at Perimeter Institute is supported in part by the Government of Canada through the Department of Innovation, Science and Economic Development and by the Province of Ontario through the Ministry of Colleges and Universities.
This work was supported partly by the Exascale Computing Project (17-SC-20-SC), a collaborative effort of the U.S. Department of Energy Office of Science and the National Nuclear Security Administration.

\end{acks}

\bibliographystyle{ACM-Reference-Format}
\bibliography{mpi,references,containers}

\appendix


\section{Handle Encoding}

This appendix presents in more detail some implementation specifics of the ABI with respect to operations, handles and datatypes.  The integer constants below are all in binary.

\subsection{Operations}
\label{app:op_handles}

The encoding for handle types is provided below:
\begin{lstlisting}
0b0000000000 invalid (uninitialized)
0b00000***** reserved handle
0b0000100000 MPI_OP_NULL
// arithmetic ops
0b0000100001 MPI_OP_SUM
0b0000100010 MPI_OP_MIN
0b0000100011 MPI_OP_MAX
0b0000100100 MPI_OP_PROD
0b00001001** reserved arithmetic op
// binary ops
0b0000101000 MPI_OP_BAND
0b0000101001 MPI_OP_BOR
0b0000101010 MPI_OP_BXOR
0b000010**** reserved bit op
// logical ops
0b0000110000 MPI_OP_LAND
0b0000110001 MPI_OP_LOR
0b0000110010 MPI_OP_LXOR
0b000011**** reserved logical op
// other ops
0b0000111000 MPI_OP_MINLOC
0b0000111001 MPI_OP_MAXLOC
0b00001110** reserved other op
0b0000111100 MPI_OP_REPLACE
0b0000111101 MPI_NO_OP
0b000011111* reserved other op
0b00******** reserved handles
\end{lstlisting}

\subsection{Other Handles}
\label{app:opaque_handles}

The encoding of opaque handles is shown below:
\begin{lstlisting}
// communicator
0b0100000000 MPI_COMM_NULL
0b0100000001 MPI_COMM_WORLD
0b0100000010 MPI_COMM_SELF
0b0100000011 reserved comm
// group
0b0100000100 MPI_GROUP_NULL
0b0100000101 MPI_GROUP_EMPTY
0b01000001** reserved group
// windows
0b0100001000 MPI_WIN_NULL
0b01000010** reserved win
// files
0b0100001100 MPI_FILE_NULL
0b01000011** reserved file
// sessions
0b0100010000 MPI_SESSION_NULL
0b010001**** reserved session
// messages
0b0100010100 MPI_MESSAGE_NULL
0b0100010101 MPI_MESSAGE_NO_PROC
0b01000101** reserved message
// error handler
0b0100011000 MPI_ERRHANDLER_NULL
0b0100011001 MPI_ERRORS_ARE_FATAL
0b0100011010 MPI_ERRORS_RETURN
0b0100011011 MPI_ERRORS_ABORT
0b01000111** reserved handle
// requests
0b0100100000 MPI_REQUEST_NULL
0b01001000** reserved request
0b01******** reserved handle
\end{lstlisting}

\subsection{Datatypes}
\label{app:datatype_handles}

Examples for datatype handles are represented below:
\begin{lstlisting}
0b1000000000 MPI_DATATYPE_NULL
// variable-size types
0b1000000001 MPI_AINT
0b1000000010 MPI_COUNT
0b1000000011 MPI_OFFSET
0b10000001** reserved datatype
0b1000000111 MPI_PACKED
0b1000001000 MPI_SHORT
0b1000001001 MPI_INT
0b1000001010 MPI_LONG
0b1000001011 MPI_LONG_LONG
0b1000001100 MPI_UNSIGNED_SHORT
0b1000001101 MPI_UNSIGNED_INT
0b1000001110 MPI_UNSIGNED_LONG
0b1000001111 MPI_UNSIGNED_LONG_LONG
0b1000010000 MPI_FLOAT
...
// fixed-size types
0b1001000000 MPI_INT8_T
0b1001000001 MPI_UINT8_T
0b1001000010 <float 8b>
0b1001000011 MPI_CHAR
0b1001000100 MPI_SIGNED_CHAR
0b1001000101 MPI_UNSIGNED_CHAR
0b1001000110 reserved datatype
0b1001000111 MPI_BYTE
0b1001001000 MPI_INT16_T
0b1001001001 MPI_UINT16_T
0b1001001010 <float 16b>
0b1001001011 <C complex 2x8b>
0b10010011** reserved datatype
0b1001001111 <C++ complex 2x8b>
0b1001010000 MPI_INT32_T
0b1001010001 MPI_UINT32_T
0b1001010010 <C float 32b>
0b1001010011 <C complex 2x16b>
...
0b1001011000 MPI_INT64_T
0b1001011001 MPI_UINT64_T
0b1001011010 <C float64>
0b1001011011 <C complex 2x32b>
...
\end{lstlisting}


\end{document}